# Patterned deposition at moving contact lines

Uwe Thiele[*]

*Department of Mathematical Sciences, Loughborough University,*

*Loughborough, Leicestershire, LE11 3TU, UK*

## Abstract

When a simple or complex liquid recedes from a smooth solid substrate it often leaves a homogeneous or structured deposit behind. In the case of a receding non-volatile pure liquid the deposit might be a liquid film or an arrangement of droplets depending on the receding speed of the meniscus and the wetting properties of the system. For complex liquids with volatile components as, e.g., polymer solutions and particle or surfactant suspensions, the deposit might be a homogeneous or structured layer of solute - with structures ranging from line patterns that can be orthogonal or parallel to the receding contact line via hexagonal or square arrangements of drops to complicated hierarchical structures. We review a number of recent experiments and modelling approaches with a particular focus on mesoscopic hydrodynamic long-wave models. The conclusion highlights open question and speculates about future developments.

[*]Electronic address: u.thiele@lboro.ac.uk; URL: http://www.uwethiele.de

## I. INTRODUCTION

Knowledge about the various interfacial effects on small scales becomes increasingly important because of the intense drive towards a further miniaturisation of fluidic systems that are used in micro- [1] and eventually nano-fluidic [2] devices. A particularly interesting example are deposition processes involving moving contact lines where physical processes on the nanometer- and micrometer scale interact in the deposition of layers of various materials (mostly but not exclusively on solid substrates). The resulting layers have macroscopic extensions, but might only be a few nanometers thick. The layers can be homogeneous or structured with lateral structure lengths that are often in the sub-micrometer or lower micrometer range.

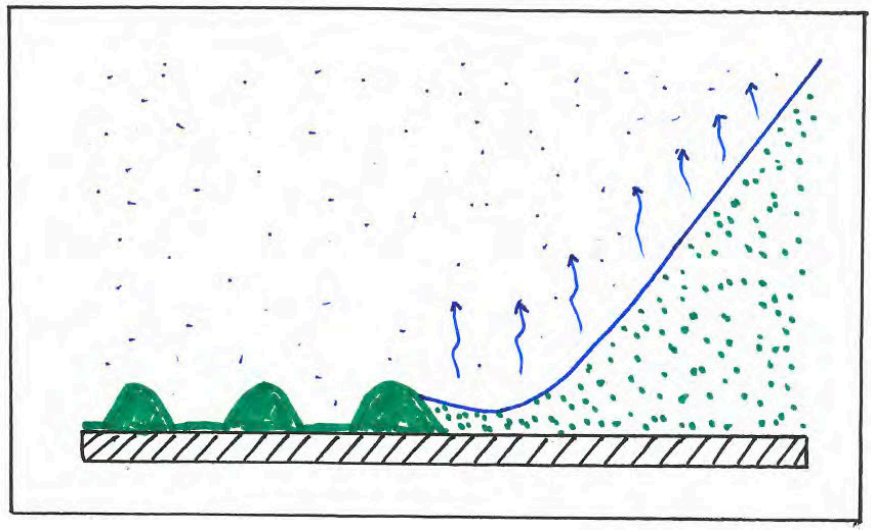

FIG. 1: Sketch of the essential core part of the geometry of every deposition process where material is left behind by a moving contact line. In the frame of the substrate the contact line region moves to the right together with the entire meniscus.

Fig. 1 sketches the typical situation close to the three-phase contact line region: In the frame of the solid substrate the three-phase contact line region – where substrate, liquid and gas phase meet – moves to the right either purely by evaporation or supported by dewetting processes or external forces. The liquid is a solution or suspension where the solute is normally non-volatile and the solvent is volatile. The solvent evaporates (often stronger close to the contact line), the local concentration of the solute increases and it is left behind.

The system is intensely investigated as on the one hand it is a practically very widely used method to deposit and structure thin layers of material on solid surfaces (see, e.g., the recent re-

Preprint– contact: u.thiele@lboro.ac.uk – www.uwethiele.de – June 29, 2013

views in Refs. [3, 4] and the introduction of Ref. [5]). Note that it is a special case of a wider class of patterning strategies that use films, drops or contact lines of solutions and suspensions with volatile solvents (see, e.g., review [6]). On the other hand the ongoing interacting non-equilibrium processes are all interesting by themselves as they are related to a number of long-standing problems in various sub-fields of hydrodynamics and soft matter science that are still under vivid discussion:

(i) *Moving contact lines* are even for simple non-volatile liquids under hot discussion. Particular keywords are the relaxation of the stress-singularity at moving contact lines, determination and prediction of dynamic contact angles, contact angle hysteresis (see reviews [7–9] and the recent Discussion and Debate volume about wetting and spreading published by the European Physical Journal Special Topics [10];

(ii) The *dynamics of the liquid-gas phase transition* at liquid-gas interfaces, i.e. the processes of evaporation and condensation, pose intriguing problems, particularly close to three-phase contact lines. See, e.g. reviews [11–13] and discussions in Refs. [14–16];

(iii) The *equilibrium and non-equilibrium phase behaviour and rheology* of high-concentration suspensions and solutions is even for bulk systems still of large present interest in soft matter science. Jamming, phase separation, gelling, crystallisation, glass transition may all occur when the concentrations reach high levels, depending on the molecular interactions of the various components. As many of these processes are even individually not fully understood, their interaction with free surfaces, moving contact lines and solvent evaporation pose challenging problems. See Refs. [17–20] as entrance points to the vast literature.

This list already indicates why experiments discover such a rich spectrum of phenomena and why it is so difficult to extract a consistent picture from the experiments and emerging models. In the present brief review, first, in section II we mention a number of experiments with a focus on the various deposition patterns found and the related quantitative measures. This is followed in section III by a brief overview of model types used in the literature and a more detailed analysis of results obtained with hydrodynamic long-wave models. Note that we will mention several treatments of evaporation in passing, but do by no means intent to review evaporation of simple liquids For recent pertinent overviews see other contributions in the present volume [?] (**citations to be introduced later in coordination with editor**), the reviews [11, 12, 14] and the introductions of Refs. [15, 21–23]. The review concludes with a number of proposals as to what are the most challenging problems and with some recommendations about set-ups that would allow us to most easily compare experimental and theoretical results.

## II. EXPERIMENTS

Deposition techniques involving a moving contact line have been studied at least since the early 20th century when Küster studied "rhythmic cristallisation" at receding contact lines of evaporating droplets of various solutions on gel substrates mentioning line patterns, zig-zag patterns, lines with side branches, flower-like arrangements of striped domains, etc. [24]. The field remained active during the following decades (see, e.g., Ref. [25]), and became also important in the context of the assembly of proteins and colloidal particles into crystals (cf. discussions and reviews of the usage of evaporating films and drops in Refs. [26–29]).

Over the previous decade, the general interest in deposition patterns has markedly increased, possibly triggered by Deegan and co-workers' detailed investigations of the "coffee-stain effect", i.e., of the deposition patterns left behind by the receding contact line of an evaporating drop of a suspension on a smooth solid substrate [30–32]. Ref. [30] reports a wide range of deposit patterns: cellular structures, single and multiple concentric rings, and fractal-like patterns (see, e.g., Fig. 2). The creation of multiple concentric rings through a stick-slip front motion of the contact line of other colloidal liquids is also described in Refs. [33, 34]. These investigations are also related to the one of Parisse and Allain of the shape changes that drops of colloidal suspension undergo when they dry [35, 36] and the creation of semiconductor nanoparticle rings through evaporative deposition [37]. Other reported structures include crack and fracture patterns [38, 39] and hierarchical patterns of obligue lines [40] (cf. Fig. 3 (a)).

Generally, evaporating a macroscopic drop of a suspension does not create a very regular concentric ring pattern in a reproducible way, but rather results in irregular patterns of rugged rings and lines [30, 33]. To produce patterns that can be employed to fabricate devices one performs the experiments on smaller scales in a somewhat more controlled way employing various small-scale geometries that confine the liquid meniscus (sphere on flat substrate, parallel plates, capillaries, etc.) as reviewed in Ref. [3]. Experiments with both, polymer solutions [41, 42, 45] and (nano)particle suspensions [46–48] result in strikingly regular line patterns with periods ranging from 1-100$\mu$m (see, e.g., Fig. 3 (b). Line patterns can be parallel or perpendicular to the receding contact line [42, 47] and are produced in a robust repeatable manner in extended regions of parameter space. Besides the lines, a variety of other patterns may also be found, including undulated

Preprint– contact: u.thiele@lboro.ac.uk – www.uwethiele.de – June 29, 2013

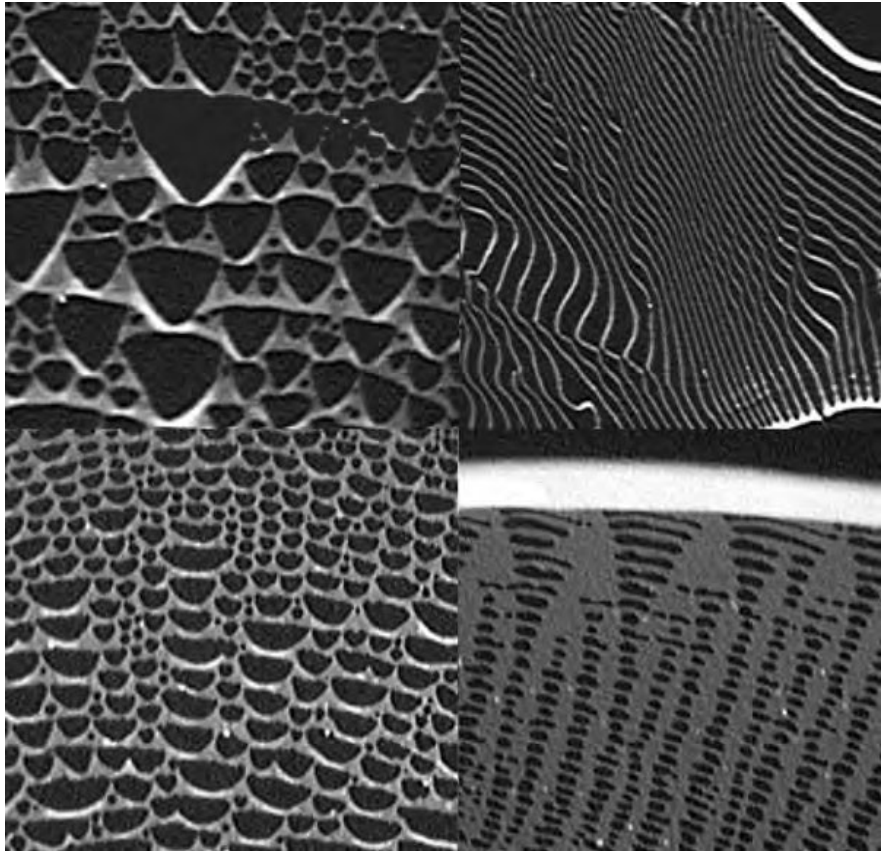

FIG. 2: Various deposits left behind by a drying drop of a suspension of 100nm polystyrene microspheres (0.5% initial volume fraction) with added anionic surfactant sodium dodecyl sulphate (SDS). The contact line moved from top to bottom. All scale bars correspond to 50 $\mu$m. In panels (a) to (d) the surfactant concentration is $8.1 \times 10^4$M, $4.3 \times 10^4$M, $1.4 \times 10^4$M, and $4.8 \times 10^5$M, respectively. Reproduced with permission from Ref. [30] (Copyright (2000) by The American Physical Society).

stripes, interconnected stripes, ladder structures, i.e. superpositions of perpendicular and parallel stripes [42] (see Fig. 3 (c)), hierarchical arrangements of pieces of parallel and perpendicular lines [43] (see Fig. 3(d)) regular arrays of drops [42, 49] or holes [41] (see Fig. 3 (e)); rings with small-scale side branches [44] (see Fig. 3(f)) and irregularly branched structures [50–53] (see review in [54]).

This type of wet evaporative deposition is now widely employed as a non-lithographic technique for covering large areas with regular arrays of small-scale structures. They are either directly deposited from the receding contact line as described above or produced using the deposited structures as templates. Examples are concentric gold rings with potential uses as resonators in advanced optical communications systems [55] and arrays of cyanine dye complex micro-domes em-ployed in photo-functional surfaces [56]. Often the patterns are robust and can be post-processed, e.g., to create double-mesh structures by crossing and stacking two ladder films [42]. A number of investigations focuses on deposition patterns resulting from more complex fluids, such as phase separating polymer mixtures [57]; solutions of the biomolecule collagen [58], liquid crystals [59], dye molecules [25, 56, 60], dendrimers [61], carbon nanotubes [62–64], DNA [65, 66], DNA and colloidal particle mixtures [67], lysozyme [68], viruses [43] and graphene [69]; and biofluids like blood [70–72]. The latter has potential medical implications as one may learn how illnesses can be detected through simple evaporation experiments on small samples [73].

Overall one finds that the deposition of regular lines is a generic phenomenon that occurs for many different combinations of substances. Examples are charge-stabilized polystyrene microspheres in water on glass [31, 74] or mica [30]; Rings were also found using metal, polyethylene, roughened Teflon, ceramic, and silicon substrates with acetone, methanol, toluene, and ethanol as solvents [32]. Used solutes are sugar and dye molecules, $10\mu$m PS spheres; 144nm PS particles in water on glass [33]; 15nm silica particles in water on glass (partially wetting, $\theta_{\mathrm{eq}} = 40$) [35, 36]; 6.5 nm silica nano-particles in water [75]; 4nm CdS particles in pyridine and 6nm CdSe/CdS core-shell particles in water on glass [37]; 90nm silica particles in pH-adjusted water on glass [46]; 0.23 $\mu$m and 3 $\mu$m poly(methyl methacrylate) (PMMA) spheres in cis-decalin (Decahydronaphthalene), and chloroform solutions of PS and poly(3-hexylthiophene) (PHT) on glass [42]; PMMA particles in octane [74], 0.1 $\mu$m and 1 $\mu$m PMMA particles in mixtures of cis- and trans-decalin [76]; bidisperse mixture of PMMA particles of different sizes in decalin [44]; benzene and chloroform solutions of PS and chloroform solution of a polyion complex on glass or mica [50]. Very similar patterns are obtained with soluble and insoluble surfactants that form monolayers on the solvent. Examples are the phospholipid dipalmitoylphosphatidylcholine (DPPC) [77–79] or poly(vinyl pyrrolidone)-coated gold nanoparticles on water [80].

To control the contact line motion various experimental setups and techniques are employed. Normally, set-ups are chosen that allow for slow evaporation. We propose to distinguish between *passive* and *active* set-ups. In the passive set-up, the solution or suspension evaporates freely and the (mean) contact line speed naturally emerges from the processes of dewetting and evaporation. In the active set-ups an additional parameter directly controls the mean contact line speed. It can often be better adjusted than the control parameters in the passive set-ups.

Examples of passive set-ups include (i) the "meniscus technique" where a meniscus with a contact line is created in a geometric confinement, e.g., in a sphere-on-flat [41, 45, 47] or ring-on-flat [26, 81] geometry, between two parallel plates [39], or in the wedge between two plates or crossed cylinders [63]; (ii) the deposition of a single drop onto a substrate where it evaporates freely [30, 33, 80]; and (iii) the deposition of flat films onto a substrate using spin-coating [82–84]. These passive set-ups are mainly controlled via the temperature, the partial pressure of the solvent, and the solute concentration. Note, however, that often they are realised at the ambient conditions of the laboratory and not in controlled environments.

Examples of active set-ups include (i) a set-up similar to blade coating where a solution is continuously provided between two glass plates while the upper plate slides backwards with a controlled velocity maintaining a meniscus-like liquid surface where the evaporation takes place and the patterns are deposited [42]. Other examples are (ii) a receding meniscus between two glass plates. Thereby the receding velocity of the meniscus is controlled by an imposed pressure gradient [48]; (iii) an evaporating drop that is pushed over a substrate at controlled velocity [46]; (iv) a solution that is spread on a substrate by a roller that moves at a defined speed [56]; and (v) a plate that is removed from a bath of the solution or suspension at a determined angle and velocity [77–80, 85]. The latter example works for the deposition of a solute from a bath of a solution. However, it also corresponds to the well-known Langmuir-Blodgett technique that is used to transfer a layer of surfactant from the free surface of a liquid bath onto a solid substrate indicating that this technique can be related to the deposition of a solute from a moving contact line. It will also turn out that the describing models are in certain limits closely related (see below in section III). For all the active set-ups it is found that additionally to the control parameters typical for the passive set-ups, the deposition patterns do also depend on the imposed mean speed at the contact line.

Up to here we have focused on experiments where the substrates are solid. However, there exist first studies of evaporating films on fluid substrates. In Ref. [86] films of a dispersion of nano-crystals in alkanes are studied that simultaneously spread and evaporate on the free surface of an immiscible polar organic fluid. As the liquid substrate is defect-free it allows for highly regular, periodic, large-area stripe patterns.

A Careful study of the rich experimental literature shows that many of the works are concerned with the creation of regular deposition patterns for particular combinations of materials in particular geometries that could be of interest for certain applications. Typical examples are shown as a proof of concept but a detailed quantitative analysis of the pattern properties in dependence of the employed control parameters is often missing, not to speak of morphological phase diagrams

Preprint– contact: u.thiele@lboro.ac.uk – www.uwethiele.de – June 29, 2013

that show which patterns are found in the various regions of the parameter space. However, without such systematic studies interactions between experiment and modelling are more cumbersome and, in consequence, an effective control of the involved processes is more difficult.

Before, modelling approaches are reviewed in section III, we give examples of quantitative analyses of experimental data. For deposition patterns that are lines parallel to the receding contact line, a typical qualitative result is the dependence of line properties like amplitude (height), period (distance), or skewness on initial concentration, imposed mean velocity of the contact line or evaporation rate. Ref. [45] shows for the sphere-on-flat geometry that the height and distance of the lines increase with the distance from the center of the concentric ring pattern. The plot reproduced in Fig. 4 also shows that height and distance of the lines also increase with the initial concentration. Different solvents are also compared quantitatively [45]. It is not ideal that the sphere-on-flat geometry (as well as the drop on flat geometry) results in a drift of parameters as during the course of the experiment the concentration in the solution often increases resulting in a drift in the characteristics of the line/ring patterns or even in a qualitative change of the pattern as the contact line moves inwards. This complicates interpretation and comparison with models.

On may say that the quantitative analysis for the Langmuir-Blodgett transfer is further advanced than for the deposition from the moving contact line of a solution or suspension. For example, Ref. [85] gives line period as a function of surface pressure and velocity of the receding plate for single species surfactant layers and mixtures of different surfactants. They also provide a first morphological phase diagrams that indicates at which parameters one finds stripes parallel to the receding contact line, stripes orthogonal to the receding contact line, and ladder structures.

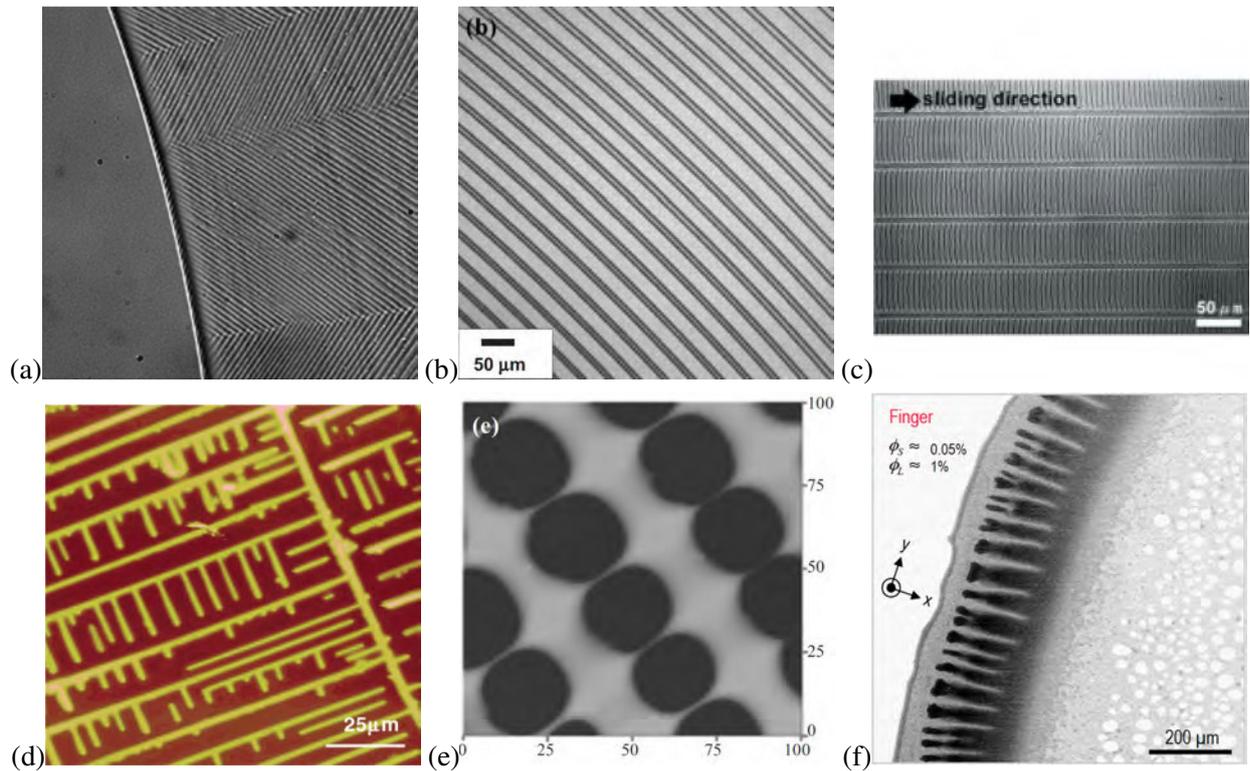

FIG. 3: Examples of various patterns obtained in drying experiments in passive geometries with various solutions and suspensions: (a) Optical image (side length 350$\mu$m) of the contact line region of a drying drop of suspension of 50 nm silica particles. The contact line recedes towards the left and leaves a hierarchical pattern of lines behind. Reproduced from Ref. [40], Copyright (2003), with permission from Elsevier; (b) Optical image zooming in on a small part of a concentric ring pattern of PMMA deposited in the sphere-on-flat geometry from a PMMA in toluene solution of concentration 0.25 mg/mL. The receding contact line was oriented parallel to the stripes. Reproduced with permission from Ref. [41], Copyright 2007 WILEY-VCH Verlag GmbH & Co. KGaA, Weinheim; (c) Optical image of ladder structures deposited in a moving cover-plate geometry from a PS in chloroform solution of concentration 4 mg/ml. The contact line receded parallel to the short lines. Reproduced with permission from Ref. [42], Copyright 2005 WILEY-VCH Verlag GmbH & Co. KGaA, Weinheim.; (d) Tapping mode AFM image of structures obtained when a solution (concentration 0.15 mg/ml) of the Cowpea Mosaic Virus (27nm size) dries on freshly cleaved mica. Reprinted with permission from Ref. [43]. Copyright (2002) American Chemical Society; (e) AFM height images (side length 100$\mu$m) of punch-hole-like PS patterns deposited from a PS toluene solution. The receding contact line was oriented parallel to the lines of holes. Reproduced with permission from Ref. [41], Copyright 2007 WILEY-VCH Verlag GmbH & Co. KGaA; (f) Confocal microscopy image of a ring-with-sidefingers structure obtained from an evaporating droplet of a bidisperse suspension of PMMA particles in decalin [44]. Reproduced with permission from Ref. [44] (Copyright (2013) by The American Physical Society).

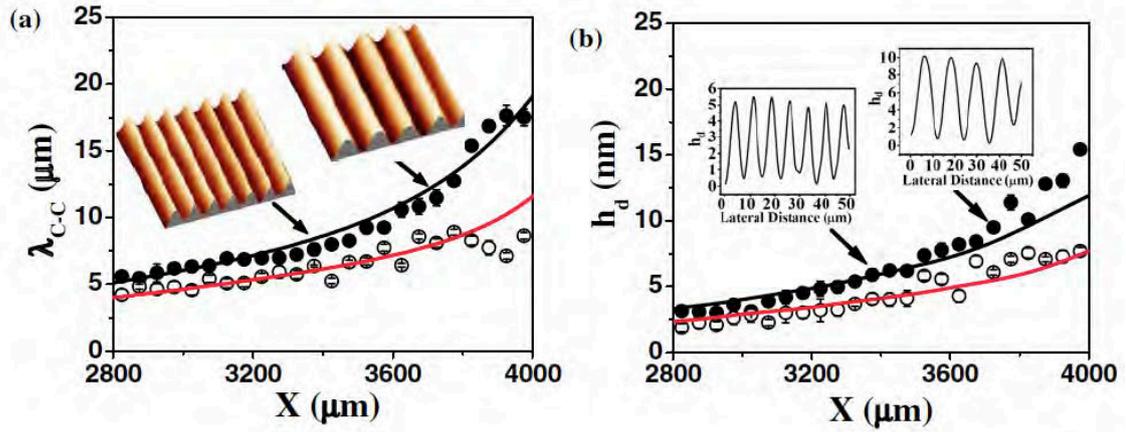

FIG. 4: (color online). Quantitative characteristics of line patterns of the polymer poly[2-methoxy-5-(2-eth-ylhexyloxy)-1,4-phenylenevinylene] (MEH-PPV) deposited in the passive sphere-on-flat geometry [45]. Solid and open circles correspond to toluene solutions with initial MEH-PPV concentrations of 0.075 mg/ml and 0.05 mg/ml, respectively. Panels (a) and (b) show the center-to-center distance of adjacent lines/rings $\lambda_{C-C}$ and the ring height $h_d$ in dependence of their distance from the sphere/Si substrate contact center, respectively. For the 0.075 mg/ml solution, typical examples of 3d AFM topographical images (50x50$\mu$ m$^2$) and corresponding cross sections are given as insets in (a) and (b), respectively. The solid lines are theoretical fits obtained as described in the main text (for more in detail see pg. 3 of Ref. [45]). Reproduced with permission from Ref. [45] (Copyright (2006) by The American Physical Society).

## III. MODELS

Despite the large number and variety of experimental works that study the creation of regular line patterns and other structures from polymer solutions and colloidal suspensions, the theoretical description and understanding of the dynamics of their formation seems still rather preliminary.

In general, most authors agree that patterns of lines that are parallel to the receding contact line result from a stick-slip motion of the contact line that is caused by pinning/depinning events [30, 45, 55, 87]. Branched structures and patterns of lines orthogonal to the contact line (the latter are sometimes called spoke patterns [44, 47, 55, 67, 88]) are thought to result from transversal instabilities of the receding contact line (sometimes called fingering instabilities) [51, 89–91].

As discussed in the introduction, a full description of the involved processes needs to account for moving contact lines, the dynamics of the liquid-gas phase transition, and the equilibrium and non-equilibrium phase behaviour and rheology of high-concentration suspensions and solutions. Many of the involved non-equilibrium processes and even the underlying equilibrium phase transitions are still under discussion and we avoid to touch the related individual issues. Instead, we first give an overview over the taken modelling approaches before discussing specific effects and results for a sub-class of models namely models based on a small gradient expansion (also called lubrication or long-wave models).

Several reduced models have been developed for the deposition process. Many of them focus on the pinning/depinning process of the contact line that is responsible for the pattern deposition and combine quasi-static considerations for droplet or meniscus shapes (e.g., assuming the liquid-gas interface always forms part of a circle/sphere), assumptions about homogeneity or a certain distribution of the evaporation flux, assumptions about the shape and density of the deposit (e.g., circular or triangular cross sections), and discuss the interaction between the contact line and the deposit that is formed, in terms of a pinning force. One example is Ref. [45] where a film thickness evolution equation in lubrication approximation (see below) is used together with assumed quasi-static expressions for the meniscus and deposit shapes to obtain average velocities of the solute moving toward the capillary edge. The obtained expressions are iteratively employed to get a best fit with experimental data (solid lines in Fig. 4. However, some details of the iterative procedure as, e.g., the calculation of pinning time and pinning force are not given in Ref. [45] making it difficult to apply the approach to other systems. The pinning force is clearly defined in Ref. [48] where a line-depositing meniscus recedes between two vertical parallel plates. There the pinning

Preprint– contact: u.thiele@lboro.ac.uk – www.uwethiele.de – June 29, 2013

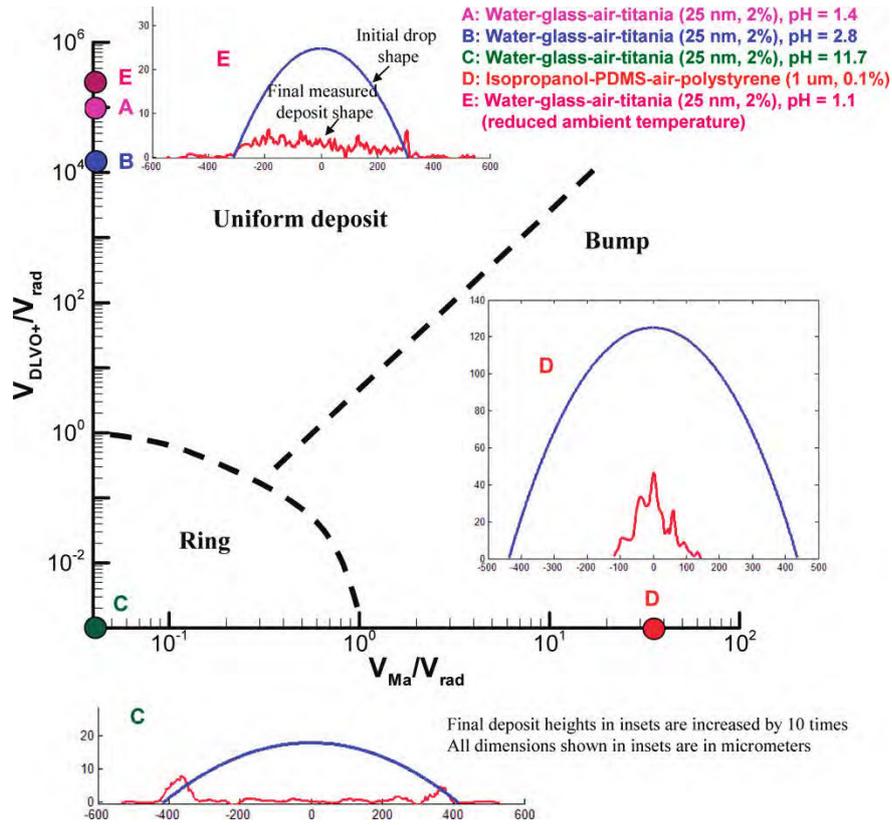

FIG. 5: (color online) Morphological phase diagramme for deposition patterns as obtained in a computational fluid dynamics approach [87, 92]. The parameter plane is spanned by ratios of the typical velocities $V_\mathrm{rad}$ of the radial flow caused by the highest evaporation rate at the pinned contact line line, $V_\mathrm{DLVO}$ related to the attractive DLVO force, and $V_\mathrm{Ma}$ related to the Marangoni flow. The letters A to D refer to experiments performed in Refs. [87, 92] while the dashed lines represent sketched boundaries of regions where homogeneous deposits, ring deposits and central deposits are expected to be found. Reprinted with permission from Refs. [92]. Copyright 2010 American Chemical Society.

force results from the difference between the equilibrium meniscus height obtained from Jurin's law (based on the balance of capillarity and gravity for a meniscus between smooth homogeneous vertical walls, see section 2.4 of Ref. [93]) and the measured rise height in the experimentally studied system where the deposits make the walls heterogeneous. The dependence of contact line pinning on colloid size and concentration in the vicinity of the contact line is investigated in Ref. [76].

Building on earlier work [31], Ref. [30] bases some estimates on the assumptions that the deposited ring is an annulus with a cross section shaped like a right triangle, the evaporating drop is always a thin spherical cap, and the volume of the drop decreases linearly. This allows for the derivation of a pair of coupled ordinary differential equations that govern the width and height of the ring deposit. The calculations in Ref. [32] assume that the drop is a spherical cap, the evaporation rapidly approaches a steady state allowing one to treat the vapour diffusion in the gas phase with the Laplace equation, i.e., by solving an equivalent electrostatic problem [31]. This results in an evaporation flux proportional to $(r_0 - r)^{-\lambda}$ where $r_0$ is the base radius of the droplet and $\lambda > 0$ depends on the contact angle. This implies that the evaporation flux diverges at the contact line and yields time dependencies of the mass of the drop, and the amount of solute arriving at the contact line that agree with experimental results [32]. The model is refined in Ref. [94] where the profile of the deposited ring is discussed. Note that these models assume that the contact line remains pinned at its initial position and are therefore not able to describe extended deposition patterns. Some of the underlying assumptions related to the diverging evaporation flux at the contact line [32] have been questioned in Refs. [22].

Ref. [74] analytically determines the flow field (including Marangoni flow) for a shallow droplet with pinned contact line and assumed spherical cap shape. The obtained velocity field is combined with Brownian dynamics simulations to study the deposition of particles modelled as simple interactionless spheres. They are convected by and diffuse in the flow. When they impact the substrate they count as deposited. The simulations show a transition from a ring-deposit (with strong Marangoni flow) to the deposition of a central bump of material (without or weak Marangoni flow).

In a computational fluid dynamics approach the system is described with a macroscale deterministic continuum model, namely, a fully non-isothermal Navier-Stokes model that consists of the complete set of transport equations for momentum, energy, and solute/colloid and vapour concentration, thereby incorporating evaporation, thermal Marangoni forces and heat transport through the solid substrate. The evaporation is limited by the vapour diffusion in the gas phase as in Refs. [22, 95–97]. Contact line motion is implemented via rules for the motion of the liquid-gas interface due to evaporation, rules for a liquid-solid transition at a critical colloid concentration, and rules for depinning when the contact angle becomes smaller than an imposed receding contact angle [87]. As a result the deposition of a single ring is modelled for a number of different parameter sets. In Ref. [92] the same authors furthermore incorporate the mesoscale elements of Derjaguin-Landau-Verwey-Overbeek (DLVO) interactions between solute particles and the solid substrate in the form of effective forces in the advection-diffusion equation for the solute concen-tration. Again, the resulting simulations show the formation of and depinning from a single ring deposit. For the parameter values used in Refs. [87, 92] no 'periodic' deposits (i.e., multiple rings) are observed. The phase diagram reproduced in Fig. 5 is proposed, where homogeneous deposits, single ring deposits and central bump deposits are distinguished that occur in different regions of the parameter plane spanned by ratios of typical velocities (see caption of Fig. 5). Although computational fluid dynamics models like the ones developed in Refs. [87, 92] contain most or all of the relevant physics, one may argue that they are tedious to use if an extensive scan of the parameter space shall be performed. Furthermore, they are rich in tricky details when it comes to incorporating wettability and contact line motion.

Alternatively, there exist modelling approaches based on microscale considerations, in particular, in the form of kinetic Monte Carlo (KMC) models for evaporatively dewetting nanoparticle suspensions [53, 84, 90, 98–100] and in the form of a dynamical density functional theory (DDFT) obtained from the KMC via coarse graining [91, 101] as (p)reviewed in Ref. [54]. Both, the microscopic discrete stochastic KMC and the continuous deterministic DDFT are able to qualitatively describe the strong fingering instability of an evaporatively receding contact line of a nanoparticle suspension and its dependence on the chemical potential of the gas phase, solute mobility and solvent-solute, solvent-solvent and solute-solute interactions. However, they do not account for convective motion of the liquid as all transport is by diffusion. These approaches could up to now not reproduce the deposition of regular line patterns parallel to the receding contact line, although line patterns orthogonal to the receding contact line result when solvent-solute decomposition is likely at the receding contact line (though not very regular, and rapidly decaying into droplets, see Fig. 12 of Ref. [91]).

A class of models that may be seen as 'lying between' macroscopic hydrodynamics and microscopic dynamical density functional theory are the so called long-wave models (sometimes also called lubrication models or small gradient models or thin film models). The subset of them that incorporates wettability via an additional pressure term (see below) and not via boundary conditions at the contact line represents mesoscopic hydrodynamic models. As explained in the conclusion they can be more easily expanded to incorporate additional physical effects like solvent-solute interactions or solute-dependent wettability than computational fluid dynamics models.

Long-wave models for the evolution of films of liquids and drops on solid substrates are derived from the full macroscopic bulk hydrodynamic equations and boundary conditions at the solid substrate and the free surface through an expansion in a small parameter, namely, the ratio of typical length scales orthogonal and parallel to the substrate. For reviews and examples of derivations see Refs. [102–104]. In the case of a partially wetting liquid, the small parameter is of the order of the equilibrium contact angle. For a drop or film of a simple volatile liquid in an isothermal situation the long-wave expansion results in the evolution equation

$$\partial_t h = -\boldsymbol{\nabla} \cdot \mathbf{J}_{\text{conv}} - J_{\text{evap}} = \boldsymbol{\nabla} \cdot [Q(h)\boldsymbol{\nabla} p] - J_{\text{evap}} \qquad (1)$$

for the film height $h(\mathbf{x}, t)$. Here, $Q(h) = h^3/3\eta$ is the mobility function in the case of a no-slip condition at the substrate where $\eta$ is the dynamic viscosity (for the case of slip see, e.g., Ref. [105]); $p = -\gamma \Delta h - \Pi(h)$ corresponds to the pressure where $\gamma$ is the liquid-gas interface tension, $-\gamma \Delta h$ is the Laplace or curvature pressure, and $\Pi(h) = -df/dh$ is the Derjaguin or disjoining pressure [8, 106, 107]; $\mathbf{x} = (x, y)^T$ and $\boldsymbol{\nabla} = (\partial_x, \partial_y)^T$. Note that in the absence of additional sources of energy the conserved part $\boldsymbol{\nabla} \cdot \mathbf{J}_{\text{conv}}$ of the r.h.s. of Eq. (1) can always be written as a gradient dynamics writing the pressure $p = \delta F[h]/\delta h$ as the variational derivative of the underlying Lyapunov functional (sometimes called effective interface Hamiltonian or surface free energy functional [9, 108, 109])

$$F[h] = \int d\mathbf{x} [\gamma \xi + f(h)] \qquad (2)$$

where $f(h)$ is the wetting energy per substrate area, and $\gamma \xi$ is the energy of the (curved) free surface per substrate area [110, 111]. Here, $\xi d\mathbf{x} \approx (1 + \frac{1}{2}|\boldsymbol{\nabla} h|^2) d\mathbf{x}$ is the surface area element in long-wave (or small-gradient) approximation. The situation is not as clear for the non-conserved part $J_{\text{evap}}$ of the dynamics in Eq. (1). Many forms are used in the literature as further discussed below.

In general, evaporation is controlled by the phase transition process at the free liquid-gas interface *and* by mass and energy transfer in the gas and liquid phase (and the substrate – for a discussion see, e.g., Ref. [112]). In consequence, one often distinguishes the limiting cases of evaporation limited by vapour diffusion in the gas phase and of evaporation limited by phase transition. In the latter case one would expect the evaporation flux to take the gradient dynamics form

$$J_{\text{evap}} = Q_{\text{nc}}(h) \left( \frac{\delta F[h]}{\delta h} - \mu \right), \qquad (3)$$

where $\mu$ is the (constant) chemical potential of the gas phase. It is (with different choices for the mobility $Q_{\text{nc}}(h)$) equivalent to evaporation fluxes used, e.g., in Refs. [5, 21, 23, 113] and accounts for the Kelvin effect (curvature influence on evaporation) and the dependence of evaporation on

Preprint– contact: u.thiele@lboro.ac.uk – www.uwethiele.de – June 29, 2013

wettability (not included in Ref. [23]). For the class of evaporation models that assume that evaporation is controlled by diffusion in the gas phase; see, e.g., Ref. [22, 97, 114] and the discussion below.

In the case of a suspension or solution the evolution equation for the film height Eq. (1) needs to be supplemented by an equation for the transport of the solute. Employing a long-wave approximation the coupled system of evolution equations for film height $h(\mathbf{x}, t)$ and height-averaged concentration $\phi(\mathbf{x}, t)$ can be readily obtained from coupled Navier-Stokes and advection diffusion equations and adequate boundary conditions [102, 104]. They are of the form

$$\partial_t h = -\boldsymbol{\nabla} \cdot \mathbf{J}_\mathrm{conv} - J_\mathrm{evap}, \tag{4}$$

$$\partial_t (\phi h) = -\boldsymbol{\nabla} \cdot (\phi \mathbf{J}_\mathrm{conv} + \mathbf{J}_\mathrm{diff}), \tag{5}$$

where most common terms in the convective and diffusive fluxes are given by

$$\mathbf{J}_\mathrm{conv} = \frac{h^3}{3\eta(\phi)} \left[ \gamma \boldsymbol{\nabla} \Delta h - \boldsymbol{\nabla} \frac{df}{dh} \right], \tag{6}$$

$$\mathbf{J}_\mathrm{diff} = -D(\phi) h \boldsymbol{\nabla} \phi. \tag{7}$$

Here, we have assumed that there is no slip at the solid substrate. Various evaporation fluxes $J_\mathrm{evap}$ are used in the literature as, e.g., the one introduced above in Eq. (3). We discuss other options below along with the various versions of long-wave models.

Such evolution equations are employed in a number of studies of drying films of solutions and of deposition processes from contact lines of solutions with volatile solvent. However, only very few studies allow contact lines to move and are therefore, in principle, able to describe the dynamics of a periodic deposition process, i.e., the stick-slip character of the process [5, 114, 116–118]. Many works focus on evaporating drops with a contact line that always remains pinned at its initial position [115, 119–121]. This implies that they are only able to describe how a deposit forms for a fixed drop base, even if fully dynamic long-wave models are employed.

An early example for such a study with pinned contact line is Ref. [119]. It uses the radially symmetric form of Eqs. (4)-(7), neglects solute diffusion ($D = 0$), assumes constant (solute-independent) viscosity ($\eta(\phi) = \eta_0$), and only accounts for the Laplace pressure term in Eq. (6) (no Derjaguin pressure, i.e., no influence of wettability). There are no further flux contributions.

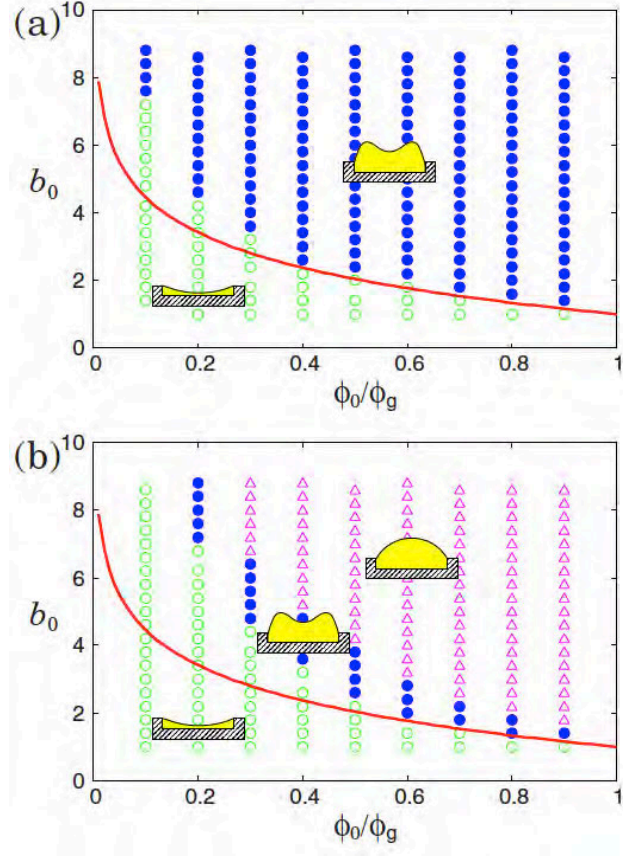

FIG. 6: (Color online) Morphological phase diagram obtained with a long-wave model that combines a quasi-static treatment of the film height profile with full time-dependent calculations of the concentration field, and also introduces a simple model for the influence of gelation. Shown is the dependence of the final deposit shape on the parameters initial concentration $\phi_0$ (in units of the gelling concentration $\phi_g$) and initial drop height $b_0$ for cases (a) without and (b) with solute diffusion. Open circles, closed circles, and triangles stand for basin type, crater type, and mound type deposits, respectively, as illustrated by the pictograms. The solid line represents a theoretical curve separating basin and crater type in the case without diffusion. Reproduced with permission from Ref. [115] (Copyright (2009) by The American Physical Society).

Results are presented for three different evaporation laws, namely

$$J^a_{\text{evap}} = \frac{E}{K+h}[1 - \exp(-A(r-r_0)^2)], \tag{8}$$

$$J^b_{\text{evap}} = \frac{E}{4h_{\max}}[1 - \tanh(A(r-r_0))], \tag{9}$$

$$J^c_{\text{evap}} = \frac{2E}{h_{\max}} \exp(-Ar^2). \tag{10}$$

They all correspond to fluxes that go to zero (or become very small) at the pinned contact line.

This shall model the effect of the growing deposit and avoid problems with singularities that arise for finite evaporation flux at a pinned contact line. In $J^a_{\text{evap}}$ the heat transfer between substrate and the free surface matters and determines the "nonequilibrium parameter $K$" [119], $r_0$ is the drop base radius, $E$ is an evaporation rate, and $A$ is related to the square of the inverse length over which the colloidal particles affect the evaporation [119]. $J^b_{\text{evap}}$ and $J^c_{\text{evap}}$ are "qualitative evaporative flux functions" developed to model particular experimental situations (for details see [119]). Depending on the evaporation flux used, ring deposition ($J^a_{\text{evap}}$ and $J^b_{\text{evap}}$) or deposition of a central bump ($J^c_{\text{evap}}$) are observed. Note, that many works that consider volatile pure liquids employ evaporation fluxes that also contain the first factor in the expression for $J^a_{\text{evap}}$ in Eq. (8) as, e.g.,

$$J^d_{\text{evap}} = \frac{E}{K+h} \qquad (11)$$

in Refs. [122] (where $K$ is said to measure "the degree of non-equilibrium at the evaporating interface"); and

$$J^e_{\text{evap}} = \frac{E}{K+h}\left(\frac{\delta F}{\delta h} - \mu\right) \qquad (12)$$

in Refs. [16, 21, 123, 124] (where $K$ is called the "kinetic parameter" [123], the "kinetic resistance number" [16] or is said "to measure the relative importance of kinetic effects at the interface" [124]). The limit were thermal aspects can be neglected by assuming that the latent heat is very small or/and the thermal conductivity is very large is obtained for $K \gg h$ (and redefining $E$). It is used, e.g., in Refs. [5, 15, 113]. Note that only $J^e_{\text{evap}}$ is a special case of the variational form given in Eq. (3), i.e., for $Q_{\text{nc}} = E/(K+h)$. From the point of view of a gradient dynamics the other given evaporation fluxes are not consistent with the energy functional underlying the respective conserved part of the evolution. Note that this is an observation only and does not imply a judgement. A gradient dynamics form as discussed above would not necessarily result in the case of evaporation limited by vapour diffusion, but might be expected in the case of evaporation limited by phase transition. But even in the latter case one might find $|\delta F/\delta h| \ll |\mu|$ and approximate $J^e_{\text{evap}}$ by $J^d_{\text{evap}}$ (redefining $E$).

Another study with pinned contact line is Ref. [115] that starts off with the same convective flux and general geometric setting as Ref. [119], but includes diffusion of the solute ($D \neq 0$) and, most importantly, distinguishes a fluid and a gel-like part of the drop. In their analysis, the authors treat the film height profile quasi-statically and approximate it by a parabolic shape with time-dependent coefficients that are calculated through ordinary differential equations coupled to the remaining evolution equation. For the evaporation flux a piece-wise function is assumed: it is constant from the center of the drop up to the distance from the center where the concentration passes the critical value for gelling. Further outside one has a gel, there is no convective flux and no evaporation, the drop shape is 'frozen'. The model can distinguish between final deposits of basin-, crater-, and mound-type. The crater-type deposits might be seen as corresponding to the deposition of a single ring. Typical obtained shapes of the dried-in deposits and morphological phase diagrams are reproduced in Fig. 6 for the cases with (top) and without (bottom) solute diffusion. Similar approaches are followed in Refs. [125] and [120] using evaporation fluxes $(n_s - n_\infty)/\sqrt{r_0^2 - r^2}$ and $\sqrt{1 + |\nabla h|^2}$, respectively, where $n_s$ is the saturated vapour density at the liquid-air interface and $n_\infty$ is the ambient vapour density away from the droplet. The proportionality of the evaporation flux on the local surface area does not seem to be consistent with long-wave approximation or a gradient dynamics form of the governing equations (before applying the quasi-static approximation).

The final study with pinned contact line we present here, is the one in Ref. [121] where time simulations of the evolution equations (4)-(7) are presented, again without solute diffusion ($D = 0$) and without wettability influence or further (e.g., Marangoni) fluxes. The work takes into account gelation close to the contact line by introducing a (i) concentration dependent viscosity $\eta(\phi) = \eta_0 \exp[S\phi/(1 - \widetilde{K}\phi)]$ (Mooney equation) and (ii) a concentration-dependent evaporation flux

$$J_{\text{evap}}^f = E \frac{1 - \phi^2}{K + h}. \tag{13}$$

Here, the concentration $\phi$ is in units of the concentration at the sol-gel transition, $S$ and $\widetilde{K}$ are fitting parameters. Overall, the obtained droplet shapes seem to match results of the simplified model with quasi-static drop profile discussed before (used, e.g., in Ref. [115]). As only crater-type deposits can be deduced from the drop profiles shown in Ref. [121] the question remains open how well the results of full and simplified (quasi-static) models match in the case of drying droplets with pinned contact line. To our knowledge no such comparison exists in the literature.

All the models that we have described in the previous paragraphs fix the drop base and are therefore not able to capture the deposition of multiple rings or of a regular line pattern in planar geometry. By fixing the drop base the contact angle is determined via the volume of the drop and no wettability effects need to be taken into account. This is different in the following models that allow for a freely receding drop edge either by introducing slip at the substrate or by employing a precursor film model (where the precursor film is either imposed 'by hand' or via specification of

Preprint– contact: u.thiele@lboro.ac.uk – www.uwethiele.de – June 29, 2013

a wetting energy $f(h)$ [cf. Eq. (2)] or of the related Derjaguin pressure $\Pi(h) = -df/dh$.

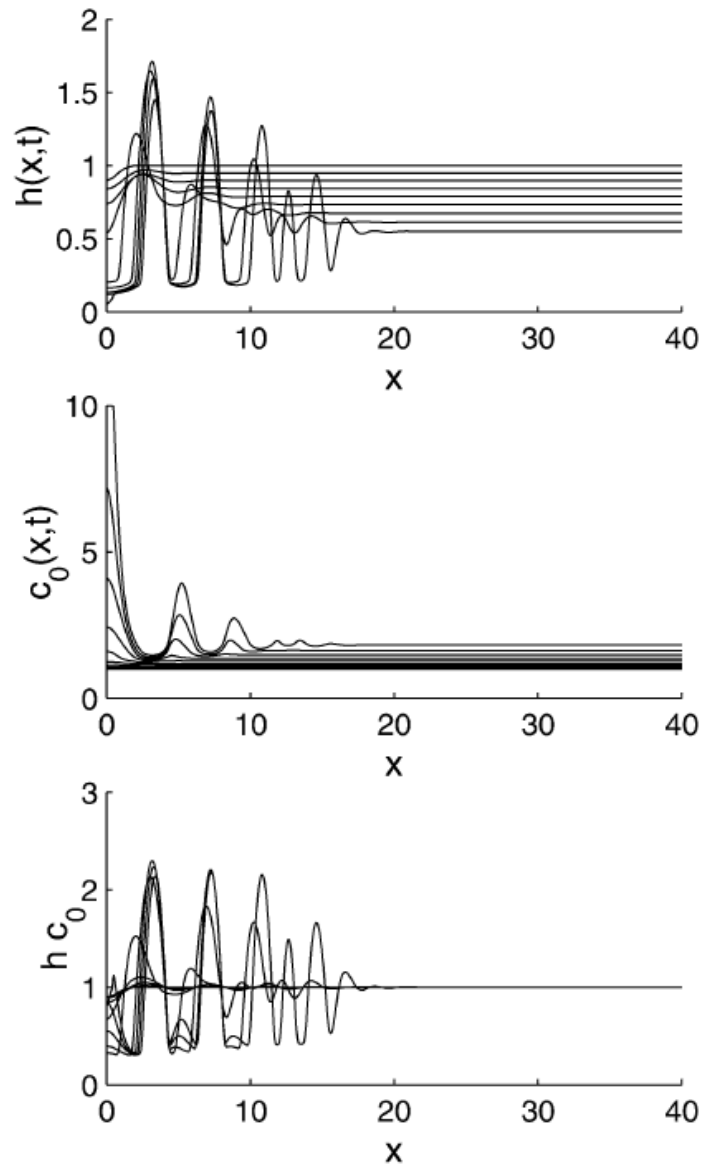

FIG. 7: Snapshots from the evolution of the film thickness (top), vertically averaged concentration (centre), and local effective solute height (bottom) obtained in a time simulation of a long-wave model for the dewetting of a suspension of non-surface-active nano-particles in a volatile solvent. For parameter values and time intervals see caption of Fig. 7 of Ref. [117]. Reprinted from Ref. [117], Copyright (2003), with permission from Elsevier.

Dewetting drying films of solutions and suspensions are studied in Refs. [116, 117] with equa-tions (4)-(7) and evaporation fluxes

$$J^g_{\text{evap}} = E_0(1-\phi)^\nu \quad (14)$$

with $0 \leq \nu < 1$, and where $E_0$ is the drying rate for the pure solvent that is assumed to be constant [116]; or corresponds to $J^d_{\text{evap}}$ (Eq. (11)) [117]. The latter work also takes vapor recoil effects into account. Both models employ concentration-dependent viscosities and concentration-independent Derjaguin pressures. For the latter they employ combinations of short-range stabilising and long-range destabilising power law contributions. Note that here we only refer to the case of surface-passive solute particles in Ref. [117] and not to the also treated case of surface-active ones. Ref. [116] investigates the dewetting and drying of an initially homogeneous film on a two-dimensional substrate with a small number of imposed wettability defects and is not directly related to deposition patterns. In contrast, Ref. [117] investigates dewetting and drying of a nanoparticle suspension on a one-dimensional substrate starting with a single initial front and observes the development of an array of drops/lines. An example is given in Fig. 7. Inspecting the figure one notes that the lines develop starting from the left where the initial position of the front is located. However, it is clear that the dried-in solute lines are not left behind by a moving front or contact line region. Instead, they result from liquid suspension drops/ridges that first develop in a directed convective dewetting process before they slowly dry in. As the film gets everywhere thinner, the process does not advance far towards the right. Such a directed dewetting process can occur via a spatially propagating spinodal process or a sequence of (secondary) nucleation processes as investigated for simple non-volatile liquids in Refs. [126–128] (1d) and Refs. [129, 130] (2d). Another thin film model that produces rings is introduced in Ref. [131], however, there the contact line is shifted 'by hand' if a certain condition is met.

The mesoscopic hydrodynamic model employed in the final part of Ref. [54] and in Refs. [5, 118] is nearly identical to the one for surface-passive solutes [117] that we just discussed. Both groups use the strongly nonlinear Krieger-Dougherty law for the viscosity [132, 133]

$$\eta(\phi) = \eta_0 \left(1 - \frac{\phi}{\phi_c}\right)^{-\nu}, \quad (15)$$

where $\eta_0$ is the dynamic viscosity of the pure solvent. Here, as before the solute concentration $\phi$ is a dimensionless volume fraction and $\phi_c$ is its value at random close packing where the viscosity diverges (for hard spheres $\phi_c = 0.63$). Various exponents $\nu$ are used in the considered long-wave models: $\nu = 2$ [117, 134], $\nu = 1.575$ [118], and various values [5]. In general, the exact value depends on the considered class of suspension. For particles that only interact via a hard-core repulsion, values between 1.4 and 3 are discussed, depending on their shape (for spherical particles $\nu = 1.575$.) [132]. Much lower values are reported for particles with net attractive interaction [17]. Depending on the physics of the transition at $\phi_c$ it may either be seen as jamming or gelation [17, 115]. Another difference between the models in Ref. [117] and Ref. [5] is the particular used Derjaguin pressure (though both model partially wetting liquids). Most importantly, the parameter region used in Refs. [5, 118] does not allow for directed spinodal dewetting (or by sequences of secondary nucleation events), but results for a volatile pure liquid in an evaporatively and convectively receding front (of constant speed) between a thick film and an ultrathin precursor film [113]. These ingredients are sufficient to model the deposition of regular and irregular line patterns from a receding front in a passive geometry. Next we briefly describe the mechanism of line deposition and give examples of typical results.

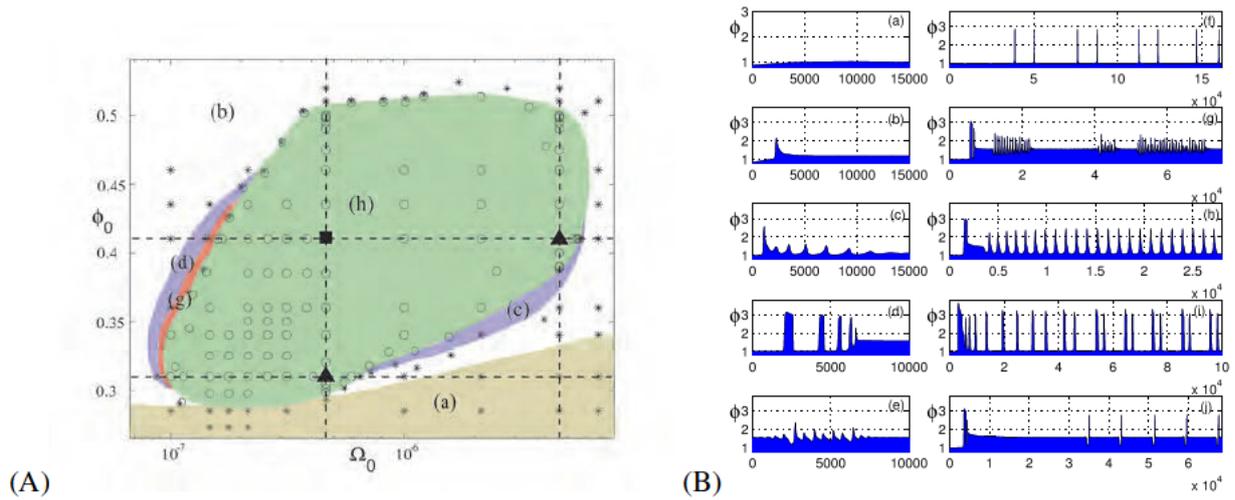

FIG. 8: (Color online) Examples of results obtained in Refs. [5, 118] with a long-wave model for the case of a passive geometry where a contact line region recedes by convection and evaporation. Panel (A) gives the morphological phase diagram of deposition patterns, in the plane spanned by the evaporation number $\Omega_0$ and the bulk concentration $\phi_0$. The letters in the differently shaded areas indicate the pattern type using the same letters as in part (B) that gives typical dried-in patterns. Most importantly, in the central area (h) regular line patterns are found after some transient (simulations denoted by ○), while outside this area (simulations denoted by ∗) a layer of constant height is deposited after a variety of transients indicated by the various shadings. In particular, one finds in (b) a single line followed by a flat layer; in (c) transient lines (whose amplitude decays first fast then slow) followed by a flat layer; in (d) transient lines (whose amplitude decays first slow then fast) followed by a flat layer; and in (f) transient double lines (converging to regular lines). In region (a) no lines are found, while in (g) one finds an intermittent line pattern [which is magnified in panel (e)]; in (h) transient lines are followed by a regular line pattern; in (i) transient lines converge to a regular pattern of double lines; and in (j) a long-period pattern switches between a flat layer and a single line with a leading depression. The corresponding parameters and further discussion can be found in Ref. [5]. Reproduced from Ref. [5] with permission from The Royal Society of Chemistry.

Refs. [5, 118] describe one of the basic mechanisms that result in the formation of regular line patterns via a self-organised cycle of pinning-depinning events, often described as a 'stick-slip' motion of the contact line. It is caused by the highly nonlinear rheology (power law divergence for suspensions in Ref. [5, 118] or exponential increase for polymer solutions in Ref. [135]): First, for sufficiently low diffusion of the solute, the ongoing evaporation rapidly increases the solute concentration in the contact line region causing a strong local increase of the viscosity. This eventually leads to a strong slow-down or even arrest of the convective motion in the contact line region. However, evaporation still moves the contact region, albeit much slower. During this phase, the material that had been collected into the contact line region is deposited as a line deposit. As the concentration in the evaporatively moving contact line region decreases, it depins from the line deposit, and moves faster again. The typical velocities in the convective and evaporative phase of motion may differ by orders of magnitude and overall the process can appear to be a stick-slip motion. Thus, the spatio-temporal self-organisation of the deposition process results from a subtle interplay of all three of the transport processes (convection, diffusion and evaporation). As even the basic model (e.g., without thermal effects, without solutal or thermal Marangoni effects) has many parameters we are still far from a complete picture.

Typical results are given in Fig. 8 where the left panel reproduces the obtained morphological phase diagram in the plane spanned by the evaporation number and the solute concentration. The right panel reproduces final dried-in patterns in the various regions of the phase diagram. Note that there is a rather extended central region of regular line patterns that are analysed in detail in Ref. [5] in their dependence on the evaporation number, solute concentration, strength of solute diffusion, wettability parameter, and viscosity exponent. This robust region of line patterns is surrounded by regions of various transient and intermittent patterns. One important part of the analysis in Ref. [5] focuses on the onset of the line patterns. Based on time simulations it was found that the regular line patterns can appear/disappear through (i) sub- or supercritical Hopf bifurcations (i.e., they disappear with a finite period and with an amplitude that reaches a (often small) finite value (sub-critical) or approaches zero (supercritical), (ii) homoclinic and sniper (saddle-node infinite period) bifurcations that are both global bifurcations [136]. In both cases the line amplitude approaches a finite value and the line period diverges when approaching the boundary - logarithmically (homoclinic bifurcation) or in a power law (sniper bifurcation). Experimentally, the subcritical Hopf- and homoclinic bifurcation may also be spotted through a hysteresis between homogeneous deposition and line deposition, while the supercritical and sniper bifurcation would not show hysteresis. Such a taxonomy of onset behaviour should proof valuable for the analysis of future experimental results - these properties are nearly not looked at until now.

In Refs. [114, 135] a thin film description of an evaporating meniscus of a suspension in an active geometry (meniscus moves under an imposed pressure gradient as in the experiments in Ref. [48]) is developed for the case of diffusion-limited evaporation. The resulting model is of the form of Eqs. (4)-(7) with an evaporation flux that is controlled by the diffusive flux of the vapour in the gas phase. The diffusive flux itself is influenced by the saturated vapour pressure at the free liquid-gas interface. In contrast to Ref. [5], Refs. [114, 135] do not take wettability effects into account (their film always remains sufficiently thick), but include a solutal Marangoni flux (only in Ref. [135]) and as well consider the dependence of the saturated vapour pressure at the free liquid-gas interface on the solvent concentration in the solution. Ref. [114] studies homogeneous deposition while Ref. [135] finds that in the considered parameter region regular line patterns are deposited in a certain range of meniscus speeds but only if a sufficiently strong solutal Marangoni effect is taken into account. From the provided numerical simulation results for the dependencies of the amplitude and period of the line patterns on meniscus speed one may conclude that at small meniscus speeds the onset is via a homoclinic bifurcation (period diverges, amplitude is finite, small hysteresis range in speed exists) and at large meniscus speeds via a subcritical Hopf bifurcation (period remains finite, amplitude is small but finite, small hysteresis range in speed exists).

Finally, we discuss dynamic models for the related process of Langmuir-Blodgett transfer where a surfactant layer is transfered from the surface of a bath onto a moving plate that is drawn out of the bath. The resulting stripe patterns are related to a first order phase transition in the surfactant layer that results from a substrate-mediated condensation effect [77]. Refs. [138, 139] develop a long-wave model consisting of coupled evolution equations for film height $h(\mathbf{x}, t)$ and

Preprint– contact: u.thiele@lboro.ac.uk – www.uwethiele.de – June 29, 2013

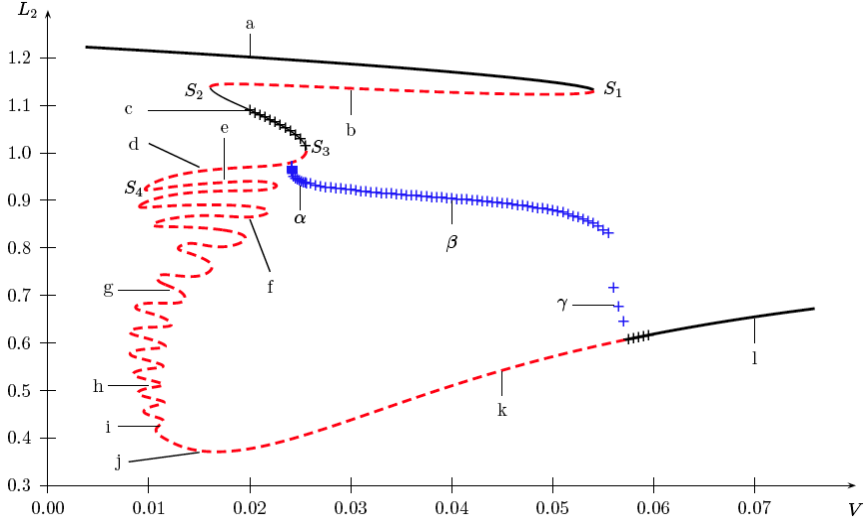

FIG. 9: Bifurcation diagram for stripe deposition obtained with a reduced (Cahn-Hilliard-type) model for Langmuir-Blodgett transfer of a surfactant layer under the influence of substrate-induced condensation. Shown is the norm of steady and time-periodic 1d concentration profiles against the velocity of the plate $V$. The solid and dashed lines represent stable (corresponding to homogeneous transfer) and unstable steady profiles, respectively, while the crosses give the time-averaged norm of time-periodic profiles that correspond to the deposition of line patterns. They emerge at low plate speed in a homoclinic bifurcation while at large plate speed several Hopf bifurcations are involved. The Greek letters label profiles given in Fig. 7 of Ref. [137] while $S_1$–$S_4$ label particular saddle-node bifurcations (see Ref. [137]). Figure reproduced from Ref. [137] under a CC BY-NC-SA licence. Copyright IOP Publishing Ltd and Deutsche Physikalische Gesellschaft.

concentration of the insoluble surfactant $\Gamma(\mathbf{x}, t)$. The general form is [102, 104]

$$\partial_t h = -\boldsymbol{\nabla} \cdot \mathbf{J}_{\mathrm{conv}} - J_{\mathrm{evap}}, \qquad (16)$$

$$\partial_t \Gamma = -\boldsymbol{\nabla} \cdot (\Gamma \mathbf{v}_{\mathrm{s}} + \mathbf{J}_{\mathrm{diff}}), \qquad (17)$$

where the evolution equation for film height is of identical form as Eq. (4). The main difference to the system (4)-(7) above is that the convective transport of the surfactant concentration $\Gamma$ is not through the film bulk flux $\mathbf{J}_{\mathrm{conv}}$ but through the liquid velocity at the free surface $\mathbf{v}_{\mathrm{s}}$. Such a model can account for the full thermodynamics of the surfactant phase transition including the resulting Marangoni fluxes [137–140]. It also contains wettability and capillarity effects. The substrate-mediated condensation is incorporated through a dependence of the free energy of the surfactant on film thickness. The model results in stripe patterns in a certain range of velocities of the moving plate [137, 138]. The stripes can be parallel (plate velocities towards upper limiting velocity) or perpendicular (plate velocities towards lower limiting velocity) to the contact line.

A disadvantage of all hydrodynamic long-wave models for line deposition is that they are still rather complicated coupled highly-nonlinear partial differential equations that do not easily lend themselves to a detailed bifurcational analysis of the patterning process. However, simulation results reveal that details of the shape of the free surface of the film and of its limited dynamics (meniscus is nearly static) do not affect the qualitative characteristics of the stripe formation process. The main role of the meniscus is to tilt the free energy potential for the surfactant phase transition from preferring one state (at hight film height) to preferring another state (at low film height). With other words, in the surfactant system the substrate-mediated condensation arises because the moving substrate drags the surfactant layer over a spatial threshold where the free energy landscape changes. This implies that the main qualitative features of the process can be captured by a strongly reduced model consisting of a Cahn-Hilliard equation (describing phase separation) [141, 142] with a double-well energy whose tilt is changed over a fixed small region in space (that represents the contact line region in the original system) [137]. The model produces stripe patterns as expected and can be analysed much more in detail than the hydrodynamic long-wave models. In particular, it is found that for the studied parameter values in the one-dimensional case the line patterns emerge at low plate velocities through a homoclinic bifurcation from an unstable branch that forms part of a family of steady heteroclinic snaking states (see bifurcation diagram in Fig. 9; for the concept of snaking cf. e.g., Ref. [143, 144]). At high plate velocities the line pattern emerges through a number of sub- and supercritical Hopf-bifurcations [137] This supports hypotheses about the onset of deposition patterns made on the basis of time simulations in Refs. [5, 137, 138]. We expect such reduced models to play an important future role in the understanding of transitions between the various two-dimensional deposition patterns.

It is interesting to note that the described onset behaviour for the deposition of regular lines may also be related to the characteristics of depinning transitions in other soft matter systems. However, to appreciate this, it is important to understand that the described transitions from homogeneous deposition to the deposition of of lines may be seen as depinning transitions in the frame moving with the mean speed of the contact line region: When a flat layer is deposited, the concentration profile is steady in the frame of the contact line region. Then, one may say that the concentration profile is pinned to the contact line region (to the moving front in Refs. [5, 118]; to the moving meniscus in Ref. [135], to the resting meniscus in Refs. [137, 138]) as it does not move relative to it. However, at the transition to periodic line deposition, the concentration profile starts to move relative to the contact line region, and one may say the concentration profile depins from the contact line region (for a detailed analysis see Ref. [5]).

This consideration makes it clear why the transition scenarios described above are very similar to the scenarios found when studying depinning in other driven soft matter systems. To illustrate how universal such transitions are, we mention two other systems: On heterogeneous substrates drops of simple nonvolatile liquids remain pinned if an external lateral driving forces remains below a threshold value. When the force passes the threshold the drops depin from the heterogeneities. Depending on system details, one finds sniper, homoclinic and super- or subcritical Hopf bifurcations [145**?**, 146]. As second system we mention clusters of interacting colloidal particles that move under the influence of external forces through a heterogeneous nanopore [147, 148]. Under weak dc driving, the particle density distribution is pinned by the heterogeneities. Depending on driving force and the attraction between the colloids, the particle density distribution may depin from the heterogeneity via Hopf and homoclinic bifurcations resulting in time periodic fluxes. The sketched comparison between the different soft matter systems shows that the emergence of regular deposition patterns (in particular, line patterns) may be related to a wider class of depinning transitions. It is to expect that particular results obtained in each system can inform future studies of the other mentioned systems.

## IV. CONCLUSIONS AND OUTLOOK

The present review has focused on deposition patterns that are left behind when a complex liquid with volatile components recedes from a solid smooth substrate. Examples are polymer, (nano-)particle and surfactant suspensions/solutions. This occurs for many combinations of substrate, solvent and solute materials in a wide range of geometries that one might classify into the two groups of passive and active set-ups. In the passive case, the evaporation proceeds freely and the (mean) contact line speed naturally emerges from the processes of convective dewetting and evaporation. In the active case, the mean contact line speed is controlled by an additional parameter as, e.g., the plate speed in the Langmuir-Blodgett transfer of a surfactant layer onto a moving plate or the pressure gradient for a meniscus that recedes between two parallel plates. The parameter that controls the mean speed of the contact line region in the active set-ups can often be better controlled than the parameters that are relevant in the passive set-ups.

First we have briefly discussed a number of examples of experimental systems. This has illustrated how rich and universal the pattern deposition process is and what the range of potential applications is. This part has also indicated that a full description of the pattern formation processes should account for moving contact lines, the dynamics of the liquid-gas phase transition, and the equilibrium and non-equilibrium phase behaviour and rheology of high-concentration suspensions and solutions – most being non-trivial non-equilibrium phenomena. The part has concluded with the assessment that although a wide range of deposition patterns is described for many systems, there is a certain lack of quantitative results that would allow us to understand how the pattern properties change with well defined control parameters, in particular, close to the onset of patterning or close to transitions between different pattern types. This is in part due to the frequent use of geometries that result in a drift of parameters during the process (as, e.g., an increasing solute concentration in a shrinking droplet). This makes a quantitative analysis very challenging and also results in modelling problems.

This has been followed by a brief overview of model types used in the literature including computational fluid dynamics, kinetic Monte Carlo simulations, dynamical density functional theory, and (mesoscopic) long-wave hydrodynamics. Subsequently, a more detailed analysis has been given of hydrodynamic long-wave models and of the results obtained with them. In this part of the review we have seen that there exists a number of long-wave hydrodynamic models for the drying of droplets with permanently fixed contact line position that mainly differ by the employed expressions for the evaporation flux and viscosity function. There are fewer works that use models which allow for freely moving contact lines. Only such models are able to describe the emergence of involved deposition patterns. Up to now they have mostly been used to analyse one-dimensional line patterns.

However, it has also become clear that the hydrodynamic long-wave models for nanoparticle suspensions and solutions are still rather restricted concerning the spectrum of physical effects that can be included in a systematic and consistent way. For instance, the models used to study line deposition do not yet account for physical effects like solvent-solute interactions or solute-dependent wettability. Also the inclusion of a solute-influence on evaporation would benefit from a more systematic approach, in particular, in the case of phase transition-controlled evaporation (for diffusion-limited evaporation see Ref. [114, 135]). A systematic way for such extensions of hydrodynamic long-wave models has recently been proposed for (i) non-surface active solutes [149, 150] and (ii) insoluble surfactants [140] improving, e.g., on earlier ad-hoc inclusions of

Preprint– contact: u.thiele@lboro.ac.uk – www.uwethiele.de – June 29, 2013

concentration-dependencies into Derjaguin pressures. For an extensive discussion see Refs. [140, 149].

To give an outlook, we sketch the main idea in the case of a non-surface active solute. For a thin film of a mixture with a neglectable influence of inertia (Stokes flow, over-damped limit) without additional sources of energy one should expect that its approach to equilibrium can be described by a gradient dynamics for the *conserved fields* film thickness $h$ and effective local solute layer thickness $\psi = h\phi$ based on an underlying free energy. Note that the non-conserved field $\phi$ is the dimensionless height-averaged per volume solute concentration. The general form of coupled evolution equations for two such conserved order parameter fields $h$ and $\psi$ in the framework of linear nonequilibrium thermodynamics is

$$\begin{aligned}\partial_t h &= \boldsymbol{\nabla} \cdot \left[ Q_{hh} \boldsymbol{\nabla} \frac{\delta F}{\delta h} + Q_{h\psi} \boldsymbol{\nabla} \frac{\delta F}{\delta \psi} \right], \\ \partial_t \psi &= \boldsymbol{\nabla} \cdot \left[ Q_{\psi h} \boldsymbol{\nabla} \frac{\delta F}{\delta h} + Q_{\psi\psi} \boldsymbol{\nabla} \frac{\delta F}{\delta \psi} \right].\end{aligned} \qquad (18)$$

The mobility matrix

$$\mathbf{Q} = \begin{pmatrix} Q_{hh} & Q_{h\psi} \\ Q_{\psi h} & Q_{\psi\psi} \end{pmatrix} = \frac{1}{3\eta} \begin{pmatrix} h^3 & h^2\psi \\ h^2\psi & h\psi^2 + 3\widetilde{D}\psi \end{pmatrix} \qquad (19)$$

is symmetric and positive definite corresponding to Onsager reciprocal relations and the condition for positive entropy production, respectively [151]. The parameter $\widetilde{D}$ is the molecular mobility of the solute. The mobility matrix is chosen in such a way that the thermodynamic form of the evolution equations (18) coincides with the hydrodynamic form (4)-(7) if the underlying free energy functional only comprises the wetting energy per substrate area $f$, the surface energy per substrate area $\gamma\xi$ and the per substrate area entropic contribution due to a low concentration solute $h\,g_{\mathrm{id}}$ (where $g_{\mathrm{id}} \sim \phi \log \phi$). That is, the free energy functional is

$$F[h, \psi] = \int \left[ \gamma\xi + f(h, \phi) + h\, g_{\mathrm{id}}(\phi) \right] dA. \qquad (20)$$

To appreciate that with this thermodynamic approach one exactly recovers the standard hydrodynamic long-wave equations (4) to (7) see Ref. [149]. With a number of small modifications this works similarly well for insoluble surfactants and in the dilute limit one recovers Eqs. (16) and (17) including the appropriate fluxes and the linear relation between surfactant concentration and surface tension [140]. However, beyond this mere reformulation the thermodynamic approach can play out its strength if one aims at the introduction of additional physical effects into the long-wave evolution equations. For instance, if instead of the purely entropic $g_{\mathrm{id}}$ one employs a double-well potential $g \sim (\phi^2 - 1)^2$ for the solvent-solute interaction (and also adds a stabilising gradient term $\sim h(|\nabla \phi|^2)$ to avoid blow-up one obtains the long-wave limit of model-H [152, 153] as derived recently via an involved long-wave asymptotic expansion [154]. One may also incorporate a concentration-dependent wettability by replacing $f(h)$ by $f(h, \phi)$. Then one obtains a concentration-dependent Derjaguin pressure as proposed, e.g., in the case of a structural Derjaguin pressure in Refs. [155–157]. This pressure, however, has to be accompanied by a flux term driven by concentration-gradients within the bulk of a thin film. This flux is neither a Marangoni nor a Korteweg flux, although it may be seen as being related to both (for details and examples in the case of a non-surface active solution see Ref. [150]). The sketched gradient dynamics approach will allow for an incorporation into mesoscopic hydrodynamics of interactions that naturally enter a dynamical density functional theory but have only incompletely been accounted for in hydrodynamic long-wave models.

Furthermore, one may add an evaporation flux to the first equation of (18). In the case of phase transition-limited evaporation, this would take a non-conserved (Allen-Cahn type) form similar to Eq. (3), i.e., it would be based on the same energy functional as the conserved dynamics in Eq. (18). However, the mobility $Q_{\mathrm{nc}}$ might depend on $h$ and $\phi$ (it may also be constant). In this way one accounts for the evaporation of a volatile solvent including the dependence of evaporation on the osmotic pressure $g - \phi g'$. To our knowledge this approach has not yet been employed in long-wave studies of deposition patterns. We also expect that the approach can be extended to other complex fluids as, e.g., suspensions of soluble surfactants or liquid crystals.

To conclude, we have given a brief overview about experimental results and modelling approaches for deposition patterns emerging at moving contact lines. Besides sketching main approaches and results, we have briefly outlined relations between the onset of patterned deposition and depinning transitions in different soft matter systems as we believe that a further analysis of such relations might be rather fruitful, in particular, as the deposition models themselves are still under development.

**Acknowledgments**

I am very grateful to the Newton Institute in Cambridge, UK, for its hospitality during my stay at the programme "Mathematical Modelling and Analysis of Complex Fluids and Active Media in Evolving Domains " and warmly acknowledge that I have benefited from many illuminating discussions with colleagues and friends. In particular, I would like to mention the other collaborators in the various "deposition projects" I was involved in, namely, A. J. Archer, L. Frastia, the late R. Friedrich, M. Galvagno, S. V. Gurevich, M. H. Köpf, H. Lopez, P. Moriarty and group, M. J. Robbins, D. V. Todorova, D. Tseluiko and I. Vancea.

As this review forms part of a volume dedicated to Manuel G. Velarde, I would also like to take the opportunity to say "Muchas Gracias" to Manuel for the inspiring time I had during my 1999/2000 stay and during later visits at the Instituto Pluridisciplinar of the Universidad Complutense de Madrid and as well for his continued interest in my work that in the course of time has benefited from our interaction in many ways.
---